\documentclass[graybox]{svmult}

\usepackage{mathptmx}        % selects Times Roman as basic font
\usepackage{helvet}          % selects Helvetica as sans-serif font
\usepackage{courier}         % selects Courier as typewriter font
\usepackage{makeidx}         % allows index generation
\usepackage{graphicx}        % standard LaTeX graphics tool
\usepackage{multicol}        % used for the two-column index
\usepackage[bottom]{footmisc}% places footnotes at page bottom

\usepackage{url}
\usepackage{booktabs}             % handles special characters in e-mail

\makeindex             % used for the subject index
\begin{document}

%\frontmatter%%%%%%%%%%%%%%%%%%%%%%%%%%%%%%%%%%%%%%%%%%%%%%%%%%%%%%

%\include{dedic}
%\include{foreword}
%\include{preface}
%\include{acknow}

%\tableofcontents
%\include{cblist}
%\include{acronym}

\mainmatter%%%%%%%%%%%%%%%%%%%%%%%%%%%%%%%%%%%%%%%%%%%%%%%%%%%%%%%
%\include{part}

%%%%%%%%%%%%%%%%%%%% author.tex %%%%%%%%%%%%%%%%%%%%%%%%%%%%%%%%%%%
%
% sample file for your "contribution" to a contributed volume
%
% Use this file as a template for your own input.
%
%%%%%%%%%%%%%%%% Springer %%%%%%%%%%%%%%%%%%%%%%%%%%%%%%%%%%

{ % begin command definition scope of contribution

\title*{Large eddy simulation of a low pressure turbine cascade with turbulent end wall boundary layers}
\titlerunning{LES of a LPT cascade with turbulent end wall boundary layers}
% Use \titlerunning{Short Title} for an abbreviated version of
% your contribution title if the original one is too long
\author{C. Morsbach \and M. Bergmann \and A. Tosun \and E. Kügeler \and M. Franke}
% Use \authorrunning{Short Title} for an abbreviated version of
% your contribution title if the original one is too long
\institute{C. Morsbach \and M. Bergmann \and A. Tosun \and E. Kügeler \at Institute of Propulsion Technology, German Aerospace Center (DLR), \email{christian.morsbach@dlr.de}
\and M. Franke \at MTU Aero Engines AG, Munich, Germany}
\maketitle

\section{Introduction}
Many large eddy simulation (LES) studies of turbomachinery flows focus on the low-pressure turbine (LPT) due to its low Reynolds number regime of $10^5$.
To further reduce the simulation costs, the assumption of a statistically 2D flow at the midspan of a turbine blade is often made by applying periodic boundary conditions in the spanwise direction.
However, a significant amount of the aerodynamic losses is generated in the secondary flow regions influenced by the interaction of end wall and blade boundary layers~\cite{Denton1993}.
Hence, the next logical step to evaluate the performance of the cascade is to conduct 3D simulations, including the endwall boundary layers, e.g.~\cite{Pichler2019}.

The MTU T161, considered in this work, is representative of high lift low-pressure turbine airfoils used in modern jet engines~\cite{Gier2007}.
The blades with a chord length of $C = 0.069935\,\mathrm{m}$ and an average aspect ratio of 2.65 are staggered at an angle of $61.72^\circ$.
The cascade is arranged with pitch to chord ratio of $t/C = 0.956$.
It features diverging end walls at an angle of $12^\circ$, such that the flow cannot be studied using a simple spanwise periodic setup.
Its geometry and boundary conditions are in the public domain and it has been the subject of both experimental~\cite{Martinstetter2010} and numerical~\cite{Mueller-Schindewolffs2017,Rasquin2021,Iyer2021} investigations.
The numerical investigations have focused on operating points with a Mach number of 0.6 and Reynolds numbers of 90,000 and 200,000 based on isentropic exit conditions.
Müller-Schindewolffs et al.~\cite{Mueller-Schindewolffs2017} performed a direct numerical simulation of a section of the profile in which the effect of the diverging end walls was modelled using inviscid walls (termed quasi 3D, Q3D).
Recently, results obtained with a second order FV code were presented, which included the end wall boundary layers but the analysis was focussed on the flow physics in the mid-section~\cite{Afshar2022}.
Various computations of the full 3D configuration were conducted using high-order codes during the EU project TILDA~\cite{Rasquin2021,Iyer2021}.
However, due to the specification of laminar end wall boundary layers and no freestream turbulence at the inflow, no satisfactory results could be obtained.
With a full 3D LES including appropriate turbulent end wall boundary layers and freestream turbulence obtained with a high order Discontinuous Galerkin (DG) method, we aim to provide a high-quality reference dataset for this configuration.

\section{Numerical method}

We use DLR's flow solver for turbomachinery applications, TRACE, to perform an implicit LES with a kinetic-energy-preserving DG scheme for the spatial discretisation of the implicitly filtered Navier-Stokes equations~\cite{Bergmann2020}.
The scheme is based on the collocated nodal Discontinuous Galerkin Spectral Element Method (DGSEM) on Legendre-Gauss-Lobatto nodes with a polynomial order of 5.
The anti-aliasing is performed by the split-formulation of Kennedy and Gruber~\cite{Kennedy2008}, cf.~\cite{Gassner2016}.
Due to the non-uniqueness of the solution at the element interfaces, Roe's approximate Riemann solver is applied for the advective part and the viscous terms are discretised by the Bassi-Rebay 1 scheme~\cite{Bassi1997}.
To advance in time, a third-order explicit Runge-Kutta scheme of~\cite{Shu1988} has been used.
Resolved turbulent scales are injected at the inflow boundary using a synthetic turbulence generation (STG) method based on randomized Fourier modes~\cite{Shur2014}.
The fluctuating velocity is introduced using a Riemann boundary condition~\cite{Leyh2020} and, in contrast to the inner faces, no numerical flux function is used at the inflow faces.

\section{Inflow boundary conditions}

The operating conditions are specified by centerline inlet flow angle of $\alpha_1 = 41^\circ$, total pressure of $p_{t, 1} = 11636 \, \mathrm{Pa}$ and total temperature of $T_{t, 1} = 303.25 \, \mathrm{K}$ as well as outlet Mach number $Ma_{2,\mathrm{th}} = 0.6$ and Reynolds number $Re_{2,\mathrm{th}} = 90,000$ based on isentropic relations computed with the centerline outflow pressure.
Furthermore it is required to match the momentum thickness of the incoming end wall boundary layers as well as the decay of freestream turbulence.
The boundary conditions are then set by a spanwise varying profiles of $p_{t, 1}(z)$ and $T_{t, 1}(z)$ with a constant $\alpha_1$ as well as spanwise varying Reynolds stress tensor $\overline{u_i' u_j'}(z)$ and turbulent length scale $L_T(z)$.
All these profiles will be obtained by scaling a boundary layer profile in wall units at $Re_\theta = 670$ (\url{https://www.mech.kth.se/~pschlatt/DATA/},~\cite{Schlatter2010}) and combining it with freestream turbulence values.
We perform a precursor channel flow simulation to determine the distance from the inlet required for the boundary layer to develop and meet the specifications.

Starting from the above centerline inflow conditions, we use the isentropic relations
\begin{equation}
    \frac{T_t(z)}{T_{1}} = \left( \frac{p_t(z)}{p_{1}} \right)^{\frac{\gamma-1}{\gamma}} = \left( 1 + \frac{\gamma - 1}{2} Ma(z)^2\right)
    \label{eqn:isentropic}
\end{equation}
to obtain a spanwise variation.
Here, we assume a constant static temperature $T_{1}$ and pressure $p_{1}$ throughout the boundary layer.
These two quantities can be computed from (\ref{eqn:isentropic}) by introducing a centerline Mach number $Ma_{1,\mathrm{center}}$, which essentially controls the value of the $T_t$ and $p_t$ at the end wall.
It has to be chosen such that the total pressure is always greater than the static pressure resulting from the simulation.
Hence, an iteration might be necessary.
It has to be noted, though, that small adaptations of $Ma_{1,\mathrm{center}}$ mainly influence the total pressure profile close to the wall and do not have a significant impact on the quantities of interest in the simulation.
For this concrete case, we chose $Ma_{1,\mathrm{center}} = 0.362$.

The discrete DNS velocity profile $u_{i,\mathrm{DNS}}^+$ and the corresponding distances to the wall $z^+_{i,\mathrm{DNS}}$ are now transformed to obtain a Mach number profile $Ma(z)$ simply using the definitions of $u^+$ and $z^+$ via
\begin{equation}
    \label{eqn:wallUnitScaling}
    %Ma(z_i) = \left. \frac{u^+_i}{u^+_N} \right|_\mathrm{DNS} \cdot Ma_{1,\mathrm{center}}, \quad z_i = z^+_\mathrm{i, DNS} u^+_{i, \mathrm{DNS}} \frac{\mu}{\rho \cdot a \cdot Ma_{1,\mathrm{center}}}
    Ma(z_i) = u^+_{i, \mathrm{DNS}} \cdot \frac{u_\tau}{a_1},
    \quad
    z_i = z^+_{i,\mathrm{DNS}} \cdot \frac{\nu_1}{u_\tau}
    \quad \mathrm{with} \quad
    u_\tau = a_1 \cdot Ma_{1,\mathrm{center}} / u_{\infty, \mathrm{DNS}}^+
\end{equation}
and the DNS freestream velocity $u^+_{\infty, \mathrm{DNS}}$.
The speed of sound $a_1$ and viscosity $\nu_1$ can be computed from $T_1$ and $p_1$ using ideal gas and Sutherland laws, respectively.
This allows to describe the spanwise variation of $T_t$ and $p_t$ using (\ref{eqn:isentropic}).
The non-dimensional turbulent stress tensor is scaled analogously via $\overline{u_j' u_k'}(z_i) = \overline{u_j' u_k'}^+|_{i,\mathrm{DNS}} \cdot u_\tau^2$.
Finally, the integral turbulent length scale in the boundary layer is estimated from the turbulent kinetic energy and dissipation rate as $L_T^+ = (k^+)^{3/2} / \epsilon^+$ and dimensionalised as the $z$-coordinate in (\ref{eqn:wallUnitScaling}).

Optimally, the freestream Reynolds stresses and turbulent length scale should be chosen such that the measured turbulent decay is matched.
However, in this case, this would result in a length scale in the order of cascade pitch.
This could, in principle, be accommodated for by simulating more than one blade at the respective expense of computational effort.
Instead, we chose to decrease the turbulent length scale and scale the non-zero components of the Reynolds stress tensor at the inflow of the domain, to reproduce the turbulence intensity in the blade leading edge plane according to the experiment.
This leads to a stronger decay of turbulent structures and has to be considered when assessing the quality of the results.

The boundary layer profile and freestream values of the turbulence quantities are combined where they intersect at the edge of the boundary layer.
Because of the large freestream turbulent length scale compared to the smaller length scale in the boundary layer, we use $L_T = \max (L_{T, \mathrm{BL}}, L_{T, \mathrm{freestream}})$ for distances to the wall greater than $\delta_{99}$.
Finally, the Reynolds stress tensor is rotated from the streamline-aligned to a Cartesian coordinate system.
Fig.~\ref{fig:inflow_and_mesh} (\textit{left}) shows both the development of momentum thickness Reynolds number $Re_\theta$ and freestream turbulence intensity $Tu$ in comparison to fits to the measured data.
As discussed, the turbulent decay is steeper in our setup but the turbulence intensity at the blade leading edge (dashed vertical line) is well captured.

\begin{figure}[t]
    \includegraphics[width=0.48\textwidth]{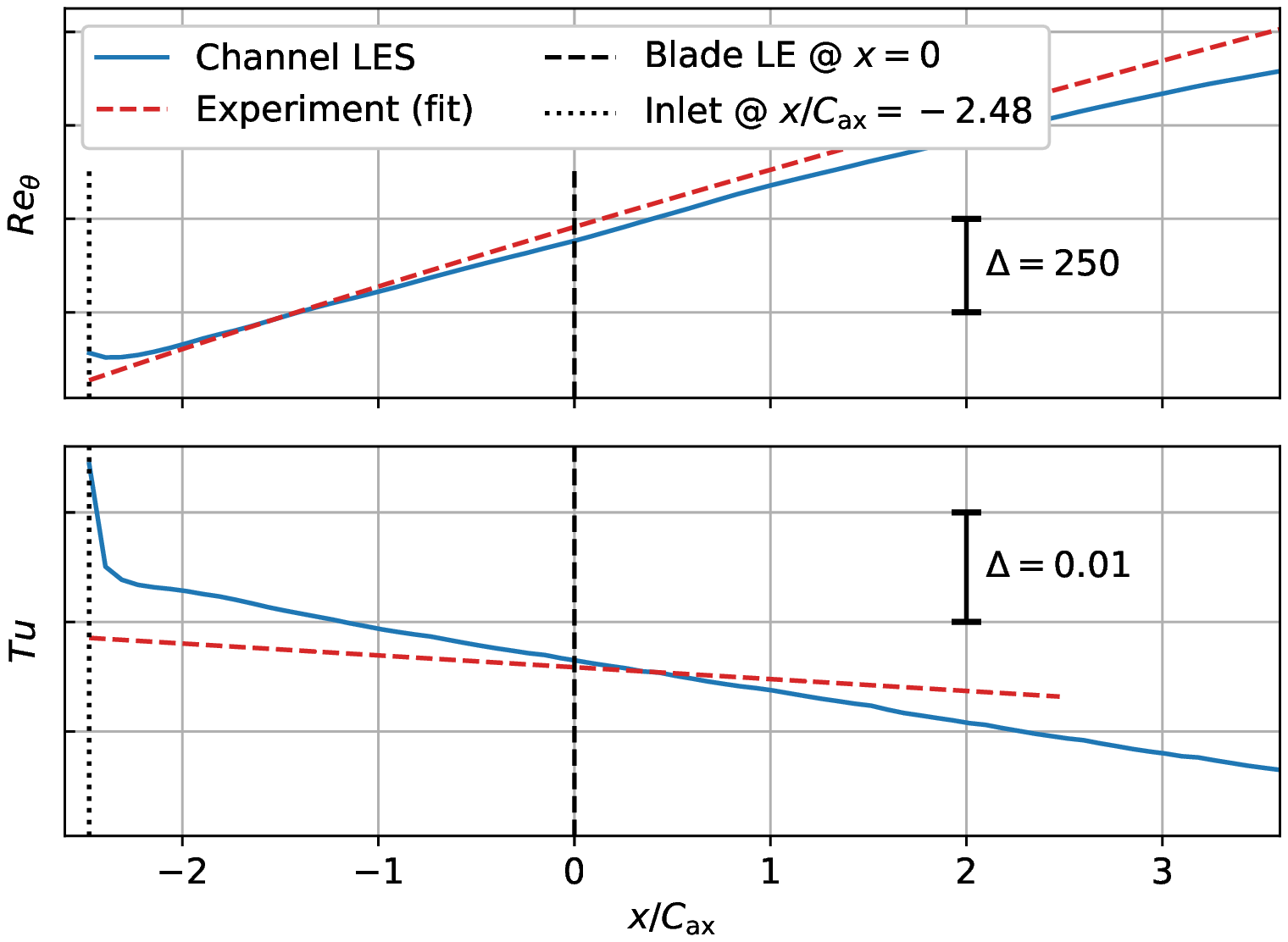}
    \includegraphics[width=0.48\textwidth]{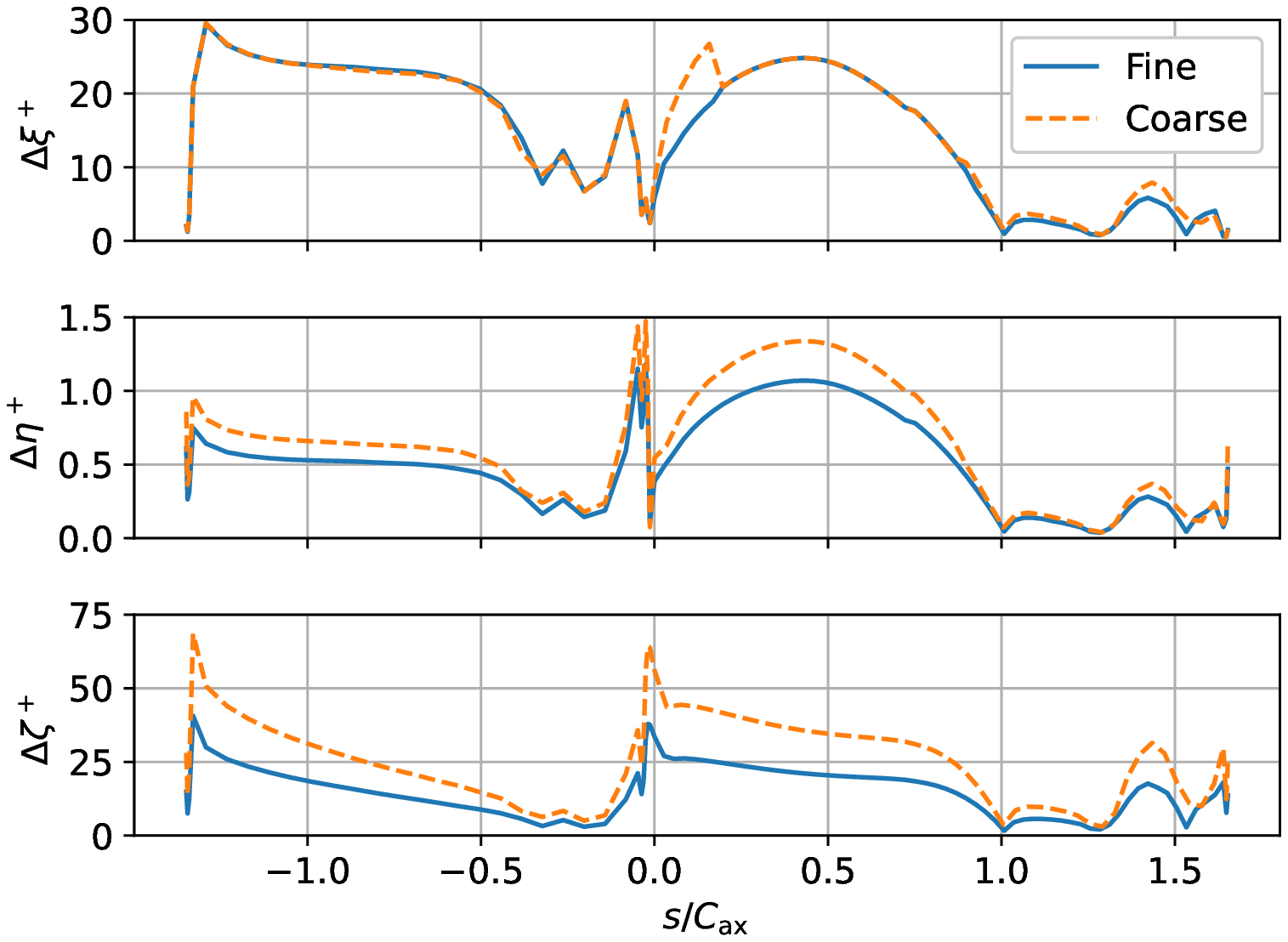}
    \caption{Momentum thickness Reynolds number and turbulence intensity over axial distance from inlet plane (\textit{left}) and average solution point distances in streamwise, wall normal and spanwise direction on the blade at midspan (\textit{right}).}
    \label{fig:inflow_and_mesh}
\end{figure}

\section{Validation and analysis of flow structures}

The mesh consisting of 876,960 hexahedral elements (geometry polynomial order $q=4$, solution polynomial order $p=5$) with 189.4M degrees of freedom (DoF) for the final computation was designed to meet widely accepted criteria for wall resolution required for LES\@.
This is demonstrated in Fig.~\ref{fig:inflow_and_mesh} (\textit{right}) which shows the streamwise, wall normal and spanwise solution point distances to be below 30, 1 and 30, respectively.
The resolution in the free stream was ensured by the ratio of solution point distance and estimated Kolmogorov scale below 6 along a mid passage streamline.
To be able to assess mesh dependence, results from a preliminary study with a significantly coarser mesh in spanwise direction with 312,424 elements ($q=4$, $p=5$) and 67.9M DoF are also included.

\begin{figure}[t]
    \includegraphics[width=0.48\textwidth]{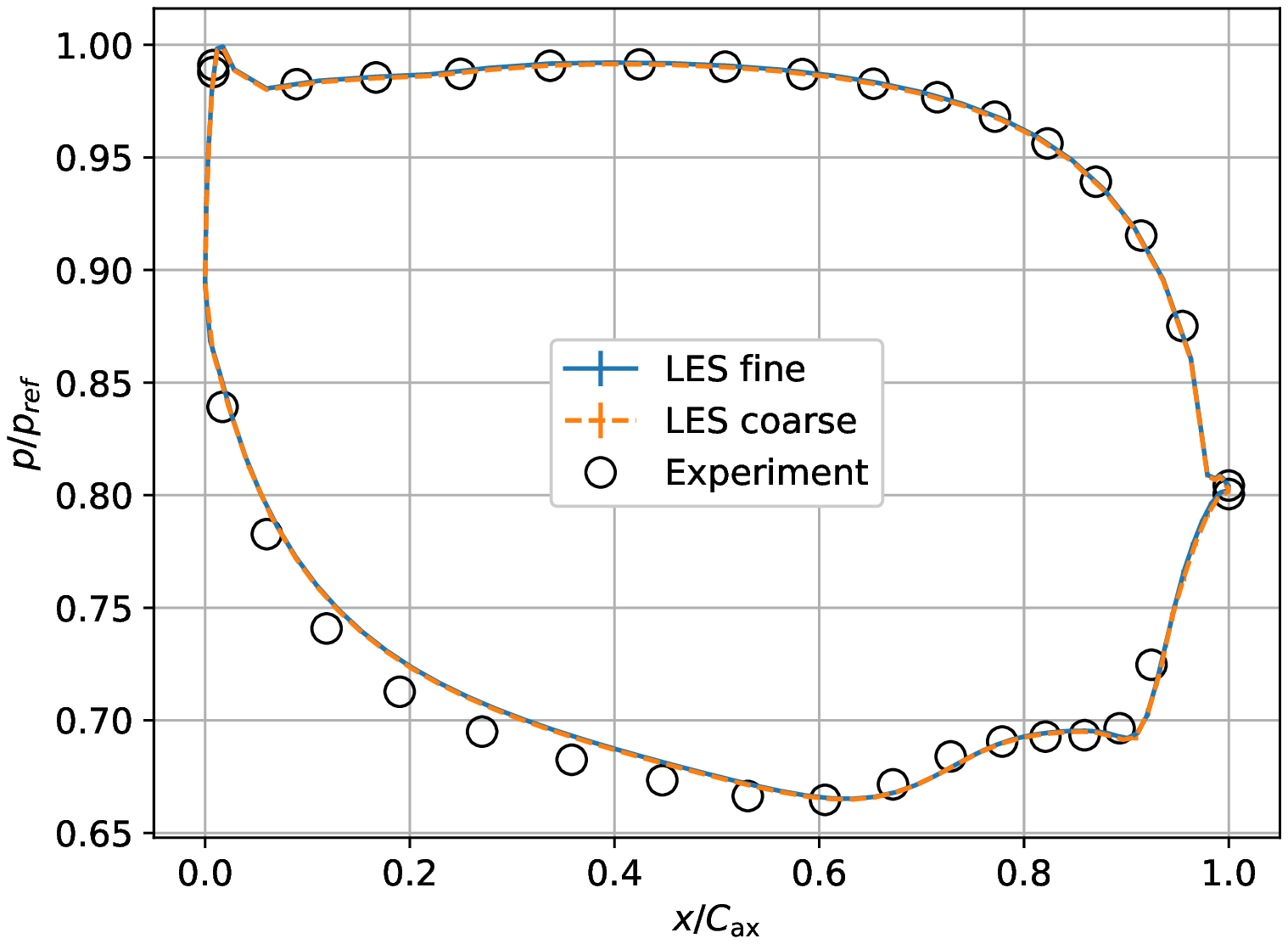}
    \includegraphics[width=0.48\textwidth]{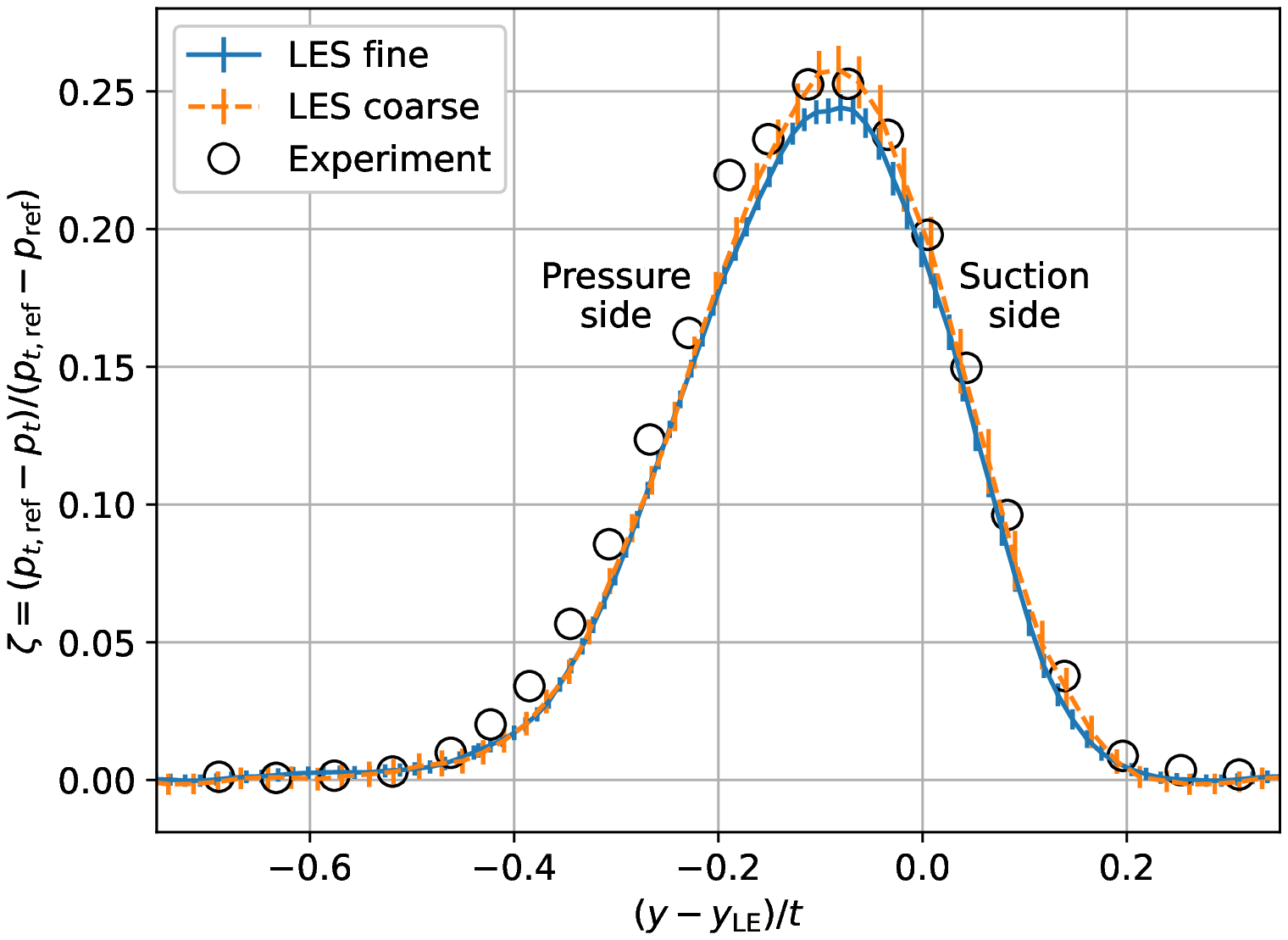}
    \caption{Midspan blade pressure distribution (\textit{left}) and wake total pressure loss coefficient at $x/c_\mathrm{ax} = 1.4$ (\textit{right}). Errorbars indicate 99\% confidence intervals.}
    \label{fig:midspan_results}
\end{figure}

\begin{figure}[t]
    % convert using: convert meanVorticesTopView.png eps3:meanVorticesTopView.eps
    \includegraphics[width=0.33\textwidth]{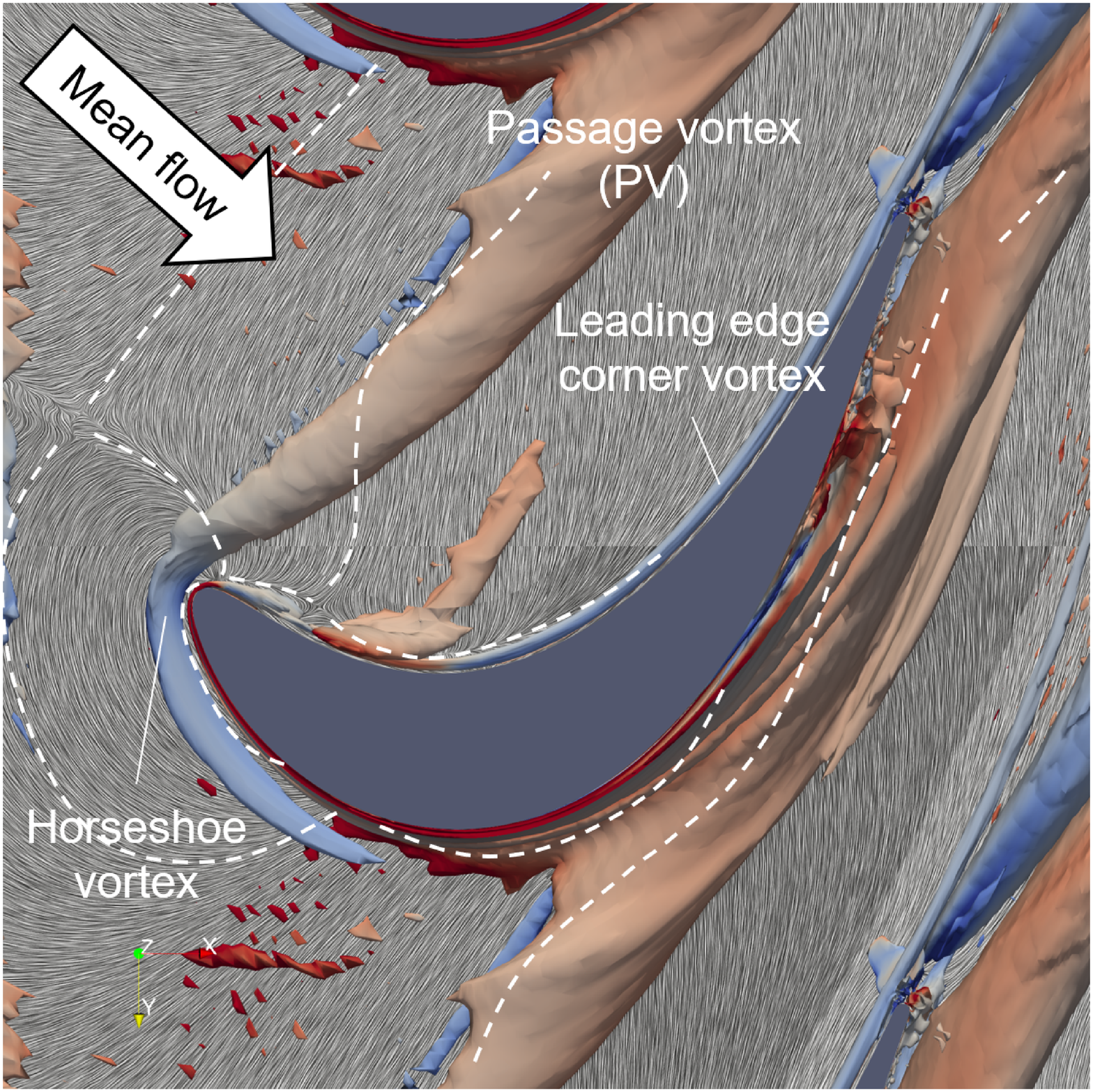}
    \includegraphics[width=0.33\textwidth]{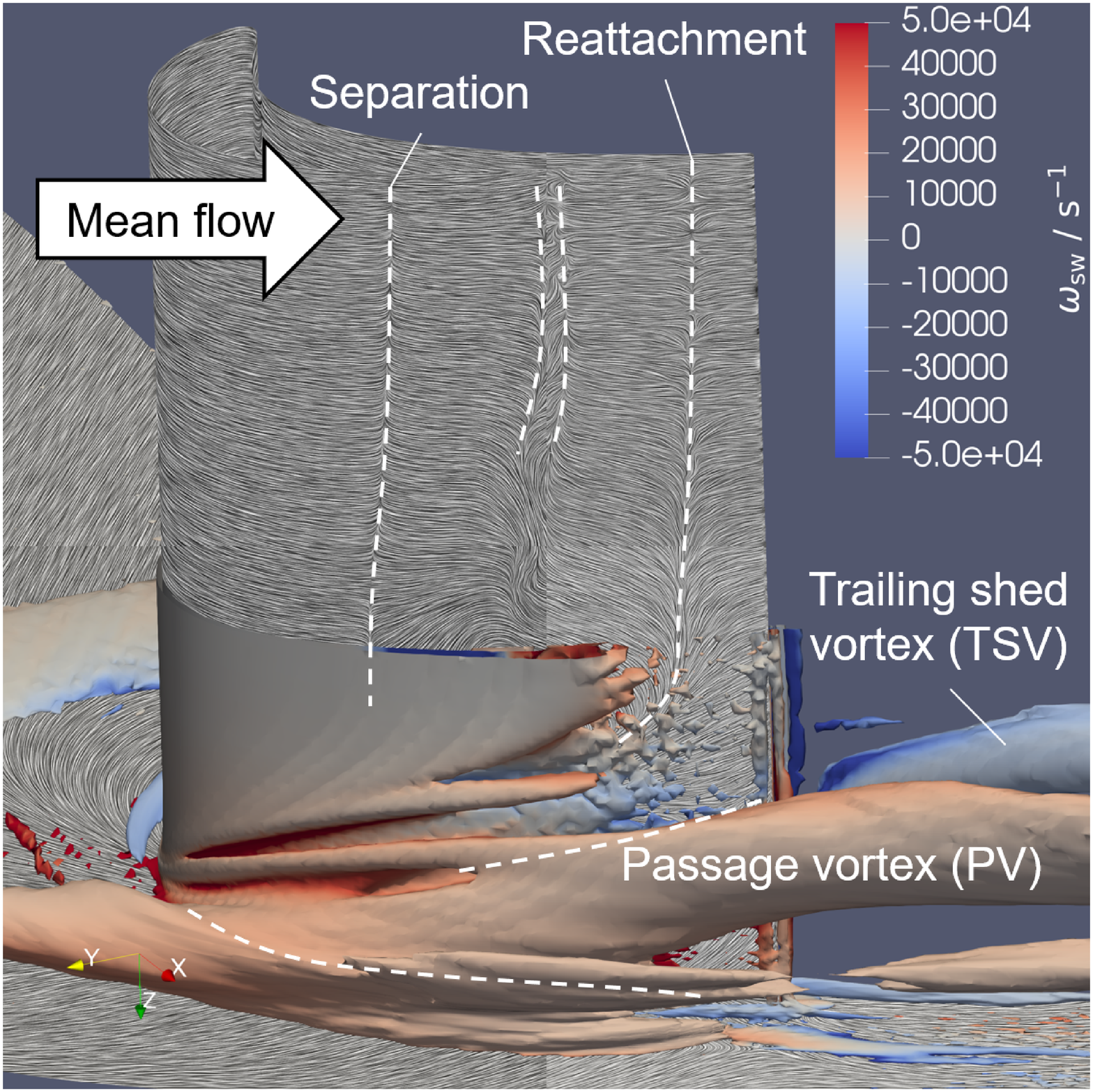}
    \includegraphics[clip, trim={0 0.5cm 0 0.5cm}, width=0.33\textwidth]{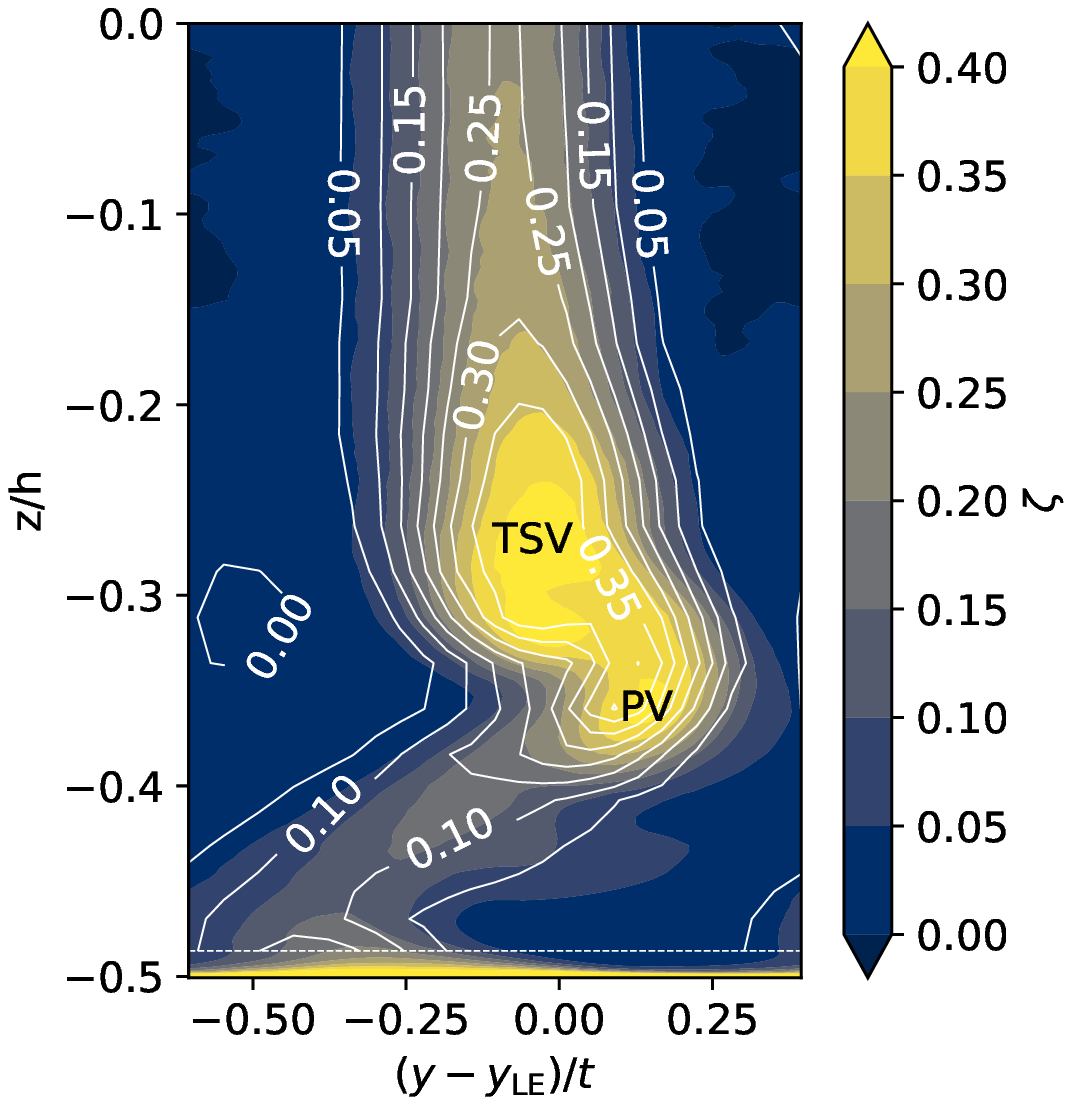}
    \caption{Mean secondary flow structures viewed towards end wall (\textit{left}) and towards suction side (\textit{middle}). Pressure loss coefficient at $x/c_\mathrm{ax} = 1.4$ (\textit{right}) - LES contour colours vs. experiment contour lines.}
    \label{fig:secondary_flows}
\end{figure}

After clearing the initial transient, the statistics on the fine mesh were sampled for 100 convective time units based on chord length and outflow velocity.
Fig.~\ref{fig:midspan_results} shows a comparison of the midspan blade surface pressure distribution and wake total pressure losses with the experiment.
Both simulations agree very well within the 99\% confidence intervals~\cite{Bergmann2021} for these first order statistics.
While LES and experiment agree very well on the pressure side, a difference in surface pressure can be found on the suction side between $x/C_\mathrm{ax} = 0.1$ and 0.6.
Similar offsets have been found in previous studies of this configuration, e.g.~\cite{Mueller-Schindewolffs2017}.
The laminar separation bubble and subsequent turbulent reattachment indicated by the pressure plateau between 0.7 and 0.9 and recovery of the base pressure, on the other hand, is captured very well.
The total pressure loss coefficient $\zeta$ is computed from time-averaged primitive variables in the wake plane at $x/C_\mathrm{ax} = 1.4$ and the upstream reference plane at $x/C_\mathrm{ax} = -0.099$.
The pitch coordinate $y$ is given with respect to the point $y_\mathrm{LE}$ on the blade at its minimum axial coordinate $x/C_\mathrm{ax} = 0$.
The simulation on the coarse mesh shows larger confidence intervals in the highly turbulent region of the wake due to its shorter averaging time of only 31 convective time units.
Both simulations and the experimental data agree very well within the statistical confidence intervals.

With the simulation setup validated against the experiment, we can now discuss the system of secondary flows developing at the intersection of blade and end wall.
It is shown in Fig.~\ref{fig:secondary_flows} (\textit{left} and \textit{middle}) in terms of time-averaged surface streaklines visualised with line integral convolution and vortices visualised using an isosurface of $Q = 10^7 \mathrm{s}^{-2}$ coloured with streamwise vorticity $\omega_\mathrm{sw}$ to indicate the sense of rotation.
Limiting surface lines have been added by hand to ease the interpretation.
The horseshoe vortex develops due to a roll-up of the incoming boundary layer.
While its suction side leg (blue) dissipates well before the suction peak, its pressure side leg follows the passage cross flow towards the suction side of the next blade lifting off the end wall and becoming the passage vortex (PV).
Behind the blade, a second large structure can be identified as the trailing shed vortex (TSV), which rotates in the opposite direction.
Both vortices persist well downstream of the blade and become visible as the two regions of strong loss in Fig.~\ref{fig:secondary_flows} (\textit{right}).
The loss distribution in the wake plane also agrees very well with the measured data.
Another small but very distinct vortex developing along the pressure side of the blade is the leading edge corner vortex.
The pressure side features a short separation bubble close to the leading edge.
Due to the diverging end walls, the backflow within the bubble is driven towards the endwalls where it rolls up, lifts off and mixes with the newly developing boundary layer in the passage between the passage vortex and the pressure side of the blade.

\section{Conclusions}

We have presented a new well-resolved dataset of the MTU T161 at $Ma_{2,\mathrm{th}} = 0.6$, $Re_{2,\mathrm{th}} = 90,000$ and $\alpha_1 = 41^\circ$ with focus on appropriate reproduction of inflow turbulent boundary layers and freestream turbulence.
Average blade loading and losses distribution due to secondary flow features agree well with the experiment underlining the validity of the presented approach.
After the brief discussion of the secondary flow structures in this paper, more in depth analysis of the unsteady data set is ongoing.
Future analyses will introduce a numerical experiment featuring unsteady wakes at engine-relevant Strouhal numbers.

}%end of command definition scope of contribution

% this is temporary, has to be replaced with the above
\bibliographystyle{aiaa}
\bibliography{literature}

\backmatter%%%%%%%%%%%%%%%%%%%%%%%%%%%%%%%%%%%%%%%%%%%%%%%%%%%%%%%
\appendix
%\include{appendix}
%\include{glossary}
%\printindex

%%%%%%%%%%%%%%%%%%%%%%%%%%%%%%%%%%%%%%%%%%%%%%%%%%%%%%%%%%%%%%%%%%%%%%

\end{document}